\documentclass[12pt,preprint]{aastex}

  \shorttitle{Solar Gamma-Ray Polarimetry}
  \shortauthors{Boggs et al.}

\begin{document}

  \title{Gamma-Ray Polarimetry of Two X-Class Solar Flares}

  \author{Steven E. Boggs\altaffilmark{1}, W. Coburn, E. Kalemci}

  \affil{
  Space Sciences Laboratory, University of California, Berkeley, CA 94720-7450
  }

  \email{boggs@ssl.berkeley.edu}

  \altaffiltext{1}{Department of Physics, University of California, Berkeley.}


  \begin{abstract}
We have performed the first polarimetry of solar flare emission
at $\gamma$-ray energies (0.2--1\,MeV). These observations
were performed with the {\it Reuven Ramaty High Energy Solar
Spectroscopic Imager} (RHESSI) for two large flares: the
GOES X4.8-class solar flare of 2002 July 23, and the X17-class
flare of 2003 October 28. We have marginal polarization detections in
both flares, at levels of 21 $\pm$ 9\% and
-11 $\pm$ 5\% respectively. These measurements significantly
constrain the levels and directions of solar flare $\gamma$-ray polarization,
and begin to probe the underlying electron distributions.
 
  \end{abstract}




  \section{Introduction}

It has been recognized for nearly four decades that X-ray
polarization could serve as a strong diagnostic
of electron beaming in solar flares \citep{kor67,elw68},
which is crucial for understanding the underlying particle 
acceleration process. While the X-ray spectral measurements
are relatively insensitive to the underlying electron
distributions, the X-ray polarization is predicted to vary
from a few percent for nearly isotropic electron distributions,
to 20--25\% for highly beamed distributions
\citep{hau72,bro72,lan77,bai78,lea83,zha95,cha96}.
The direction of the polarization vector -- whether
parallel or perpendicular to the projected magnetic field --
will depend on whether the electrons are primarily beamed along
the magnetic field lines (small pitch angle) or perpendicular
(large pitch angle).
In addition, the degree and direction of polarization
will be a function of the
the viewing angle to the flare, i.e. radial
position on the solar disk. Therefore, in order
to constrain electron beaming models, we would
ultimately like to measure the degree and direction of linear polarization
for many flares, covering the full range of viewing angles over the
solar disk.

Solar X-ray polarization observations have historically proved
difficult [see \cite{cha88,mcc04} for overviews]. 
All of these measurements 
were performed at energies below 20\,keV \citep{tin70,nak74,tin76,tra84},
where the emission is strongly contaminated by thermal emission,
significantly complicating interpretation of the measurements.
These observations and subsequent analyses have lead to the
general conclusion that the best energy range for studying
flare polarization is $>$50\,keV \citep{cha88}. 
At energies above 1\,MeV, flare emission is dominated by nuclear
lines which are expected to be unpolarized; therefore, the best
energy band for studying solar flare polarization
is 50\,keV -- 1\,MeV \citep{lei97,cha88}. To date, there
have been no attempts to study the $\gamma$-ray polarization,
and theoretical predictions above 200\,keV are limited \citep{hau72,lan77,bai78}.
Studies are currently being performed in the 50--100\,keV
band with RHESSI using an extension of the
techniques used previously in the lower energy
observations \citep{mcc02}, with results
in preparation (McConnell, private communication). This paper
is the first attempt to measure solar flare polarization
above 200\,keV, where the thermal emission is truly negligible.

While not primarily designed as a polarimeter, RHESSI is
sensitive to polarization in both the 20--100\,keV band, and
the 0.2--1\,MeV band. A passive Be block was
added to the RHESSI focal plane in order to detect polarization
in the hard X-ray range (20--100\,keV) \citep{mcc02}. At higher
energies ($>$ 200\,keV), Compton scattering of photons between
the detectors themselves can be used to measure polarization.
This technique has been used with RHESSI data to search for
polarization from $\gamma$-ray burst GRB 021206 \citep{cob03,wig04}.
While the results for this GRB remain controversial, the
work on this burst confirms the sensitivity of RHESSI to polarization
for bright $\gamma$-ray events. The factors that
made GRB 021206 so difficult to analyze (short duration, rapid
variability, off-axis location, relatively small number of counts,
telemetry decimation)
go away for these solar flares, simplifying the
analysis. Indeed, given the large number of flare photons, we have
implemented an analysis method with much stricter data selection than implemented
in our analysis of GRB 021206 \citep{cob03} in order to address
the concerns raised by subsequent analysis \citep{wig04}.

We will first give an overview of the RHESSI instrument and discuss
how it can be used to measure $\gamma$-ray polarization. Next, we discuss
in detail our methods used for this analysis. Finally, we present a
brief discussion of the scientific implications of these observations.

\section{RHESSI Instrument}

RHESSI has an array of nine coaxial germanium detectors,
designed to perform detailed spectroscopic
imaging of X-ray and $\gamma$-ray emission (3\,keV -- 17\,MeV) from
solar flares \citep{lin02}. RHESSI imaging is
performed by two arrays of opaque 1-D grids, separated by 1.5 m, and co-aligned
with the nine detectors \citep{zeh03}. As the RHESSI spacecraft rotates (4-s
period, axis aligned with the Sun) these grids modulate the count rate in
the detectors, allowing imaging through rotational modulation collimator techniques
\citep{hur02}. Thus, RHESSI has high
angular resolution (2.6$\,^{\prime\prime}$) in the 1$^\circ$ field of view of its optics.
In addition, RHESSI sends down the energy and
timing
information for each photon, allowing detailed timing measurements.

\subsection{RHESSI Spectrometer}

The heart of the RHESSI instrument is the array of nine coaxial germanium detectors,
7.1-cm diameter, 8.5-cm height each \citep{smi02}. The detectors are arranged
in the spectrometer as shown in Fig.\,1.
RHESSI performs its imaging by photon timing, not positioning, so there is
no spatial information for interactions within a detector segment -- just
energy and timing information.
The detectors are designed to be
electrically ``segmented'' into two monolithic sections,
so that the front segments ($\sim$1.5-cm thickness)
perform as separate detectors, with separate electronics, from the rear
segments ($\sim$7.0-cm thick).
RHESSI detectors are segmented in order to optimize
performance over its broad energy band:
solar X-ray and hard X-ray photons ($<$100\,keV) will
preferentially stop in the ``fronts'',
while $\gamma$-rays ($>$100\,keV) will preferentially
interact in the ``rears''.
Given the overwhelming X-ray and hard X-ray fluxes
from large flares, the fronts both measure the low energy emission, as well
as shield the rears from these overwhelming count rates, for minimal
deadtime to the $\gamma$-ray emission.
Detector \#2 initially failed to segment after
the launch of RHESSI, and therefore acts as a single detector covering
the whole front and rear volume at the times of these observations.
The fronts have a typical threshold of 2.7\,keV, and a spectral resolution
of roughly 1\,keV at 94 keV.
The rears have a typical threshold of 20\,keV, and a spectral resolution
of 3\,keV at 1.117\,MeV. When an interaction occurs in a rear segment,
there is a deadtime of 8--9\,$\mu$s before another interaction
can be measured in the same rear.

For our solar flare polarization analysis, we are using only the rears
in order to avoid the overwhelming flux in the fronts.
By including the front segments during these flares,
our livetime to real scatter events goes nearly to zero, and our
coincidence events are dominated by chance coincidences. 
In this analysis the fronts are treated effectively as passive
material.
For these same reasons, we exclude detector \#2 from this
analysis. Our polarization analysis utilizes the rear
segments of detectors \#[1,3,4,5,6,7,8,9]. In order to maximize the
signal-to-noise, we include only scatters that occur between adjacent
detector pairs since
these will dominate the real scatter events, and minimize chance
coincidences. The possible scatter paths included in our analysis
are sketched in Fig.\,1.

\subsection{RHESSI Data}

RHESSI sends down data for each individual interaction in
its detectors (``photon-mode''). Here we are careful to
differentiate ``interactions'' from ``photons.'' For
most events in the RHESSI instrument these are one and
the same. We are interested, however, in the subset of
photon events which have interactions in multiple
detectors. We have tried to be consistent here in our
use of these two terms. For each interaction,
there are three primary pieces of information: interaction time,
interaction energy, and detector segment identification.
Each interaction is tagged with a time resolution of 1 ``binary
microsecond'' (1\,b$\mu$s = 2$^{-20}$\,s), and $\sim$0.3-keV
energy sampling.

RHESSI does not have detector-detector
coincidence electronics, so coincidences have to
be determined by comparing the times of individual interactions.
Whether any
given interaction is a photon that interacted in a single detector, or
is part of a coincident photon scattering event, must
be determined by analysis of the interaction times. Between
the rear segments we use in this analysis, we
require coincidences to have $\Delta$t = 0\,b$\mu$s (i.e., within one
RHESSI time resolution unit).
This criteria is discussed in detail below.

RHESSI often enters a ``decimation''
mode where only a fraction of the events in the rears will be stored for
energies below $\sim$380\,keV, typically $\frac{1}{6}$ to $\frac{1}{4}$,
which is designed to save
on-board memory during periods of high background.
This decimation mode can significantly complicate polarization
analysis \citep{wig04}, so we have chosen only flare periods and
background periods when RHESSI was not decimating in the rears.

\subsection{RHESSI $\gamma$-ray Polarimetry}

RHESSI is not designed to be a Compton $\gamma$-ray polarimeter; however, several
aspects of its design make it sensitive to polarization.
In the $\gamma$-ray range of 0.15--10\,MeV, the dominant photon interaction in the 
RHESSI detectors is Compton scattering --- yet most photons are eventually
photabsorbed in the same detector in which they initially scattered.
A small fraction of incident photons will 
undergo a single scatter in one detector before being scattered and/or
photoabsorbed in a 
second separate detector. These scattered-photon events are
sensitive to the incident $\gamma$-ray 
polarization since linearly polarized $\gamma$-rays preferentially scatter
in azimuthal directions perpendicular 
to their polarization vector. In RHESSI, this scattering property can be used to measure 
the intrinsic polarization of solar flares.

The sensitivity of a Compton polarimeter 
is determined by its effective area to scatter events, and the average value 
of the polarimetric modulation factor,
$\mu$, which is the maximum variation in 
azimuthal scattering probability for polarized photons \citep{nov75,lei97}.
This factor is given by:
  \begin{eqnarray}
  \mu(\theta, E_{\gamma}) = \frac{d\sigma_{\perp}-d\sigma_{\parallel}}{d\sigma_{\perp}+d\sigma_{\parallel}},
  \end{eqnarray}
where $d\sigma_{\perp}$, $d\sigma_{\parallel}$ are the Klein-Nishina differential
cross sections for 
Compton scattering perpendicular and parallel to the polarization direction, 
respectively, which is a function of the incident photon energy $E_{\gamma}$,
and the Compton 
scatter angle $\theta$ between the incident-photon
direction and the scattered-photon direction. 
For a source of count rate $S$ and fractional polarization $\Pi_s$, the expected azimuthal 
scatter angle distribution (ASAD) is given by:
  \begin{eqnarray}
  \frac{\partial S}{\partial \phi} = (\frac{S}{2\pi}) [1-\mu_m \Pi_s \cos{2(\phi - \eta)}] ,
  \end{eqnarray}
where $\phi$ is the 
azimuthal scatter angle, $\eta$ is the direction of the
polarization vector, and $\mu_m$ is the 
average value of the polarimetric modulation factor for the instrument. While RHESSI 
has a small effective area for events that scatter between detectors due
to its large-volume detectors (photons have a small probability of escaping a
detector after their first scatter), it has a 
relatively large modulation factor in the 0.2--1\,MeV range, $\mu_m \sim 0.32$,
as determined by Monte Carlo simulations described below.

RHESSI also has the advantage of rotating
around its focal axis (centered on the Sun) with a 4-s period. 
Rotation averages out the effects of scattering
asymmetries in the detectors and passive materials 
that could be mistaken for a modulation. Sources that vary on timescales longer than
the rotation period will be relatively insensitive to 
the systematic uncertainties that typically plague polarization measurements.
The photon scatters between any pair of detectors will
modulate twice per rotation due to a polarized component.
We note that this modulation is orders of magnitude
slower than modulations introduced by the imaging grids \citep{hur02}.

Finally, while the RHESSI detectors have no positioning sensitivity, they are 
relatively loosely grouped on the spacecraft, allowing the azimuthal angle for a given 
scatter to be determined to within $\Delta \phi \simeq$  13$^{\circ}$ r.m.s.,
as determined by the diameter of the detectors and their typical spacing (Fig.\,1).
This angular uncertainty will decrease 
potential modulations by a factor of 0.95, which is included in our calculated 
modulation factor.

\section{Observations}

For this analysis we chose two of the brightest $\gamma$-ray flares seen by
RHESSI, both with strong emission above 0.2\,MeV. The 2002 July 23 flare was chosen
specifically due to the large volume of RHESSI imaging and spectroscopy analysis
performed on this flare. The 2003 October 28 flare was chosen both due to its
strong $\gamma$-ray emission, and because it is one of the few $\gamma$-ray
flares RHESSI has observed toward the center of the solar disk, providing a
smaller viewing angle relative to solar vertical
than the 2002 July 23 flare. For a given flare we had
two criteria for the time interval analyzed, both designed to maximize
signal-to-noise. First, we required the 0.2--1\,MeV flare photon rate to be greater than
the background photon rate. Second, we required the rate of real photon-scatter
coincidences to be larger than the chance coincidence rate (Sec.\,4.5).

\subsection{23 July 2002}
On 23 July 2002 an X4.8-class flare was observed by RHESSI through
its initial rise (00:18--00:27 UT), impulsive phase (00:27--00:43 UT),
and much of its subsequent decay [\cite{lin03}, and references
therein]. The uncorrected 0.2--1\,MeV lightcurve for the RHESSI
rear detectors is shown in Fig.\,2a. To be clear, these are the single
interactions in the rear segments. From the lightcurve
the period when RHESSI is decimating in the rears is obvious.
For this analysis, we selected the time interval UT 00:27:20--00:32:20
to study for polarization. This interval avoids decimation periods,
as well as selects a period when the flare dominates the count
rate. The emission from this flare originated at S13$^{\circ}$, E72$^{\circ}$,
near the limb of the solar disk, resulting in a viewing angle of 73$^{\circ}$.

\subsection{28 October 2003}
On 28 October 2003 the tail of the impulsive phase of an X17-class flare
was observed by RHESSI at 11:06 UT, just as the satellite was emerging from the South
Atlantic Anomaly (SAA), through which the detectors are turned off.
Therefore, RHESSI missed much of the initial rise of this flare.
In addition, the decay phase was cut off due to Earth occultation of
the sun during the RHESSI orbit. Despite these limitations,
which have minimized the standard RHESSI analysis of this flare
compared to the 23 July 2002 flare, we chose this flare for polarization
analysis for two reasons. First, it is very bright in the $\gamma$-ray
range. Second, this is the strongest flare RHESSI has seen near the
center of the solar disk (S17$^{\circ}$, E9$^{\circ}$),
with a viewing angle of 19$^{\circ}$.
The uncorrected 0.2--1\,MeV lightcurve for the RHESSI
rear detectors is shown in Fig.\,2b.
For this analysis, we selected the time interval UT 11:10:22--11:18:22
to study for polarization. This is offset from the peak measured $\gamma$-ray
rates by approximately two minutes; however, during the peak
of the emission the count rate is high enough that the livetime for
measuring scattered photons in the rear segments is near zero (Sec.\,4.5).

\section{Analysis Method}

The ultimate goal of this analysis is to study the azimuthal scatter angle
distribution (ASAD) for photons that scatter between RHESSI detectors,
as this distribution is projected {\it on the solar disk}, i.e. corrected
for the rotation of the spacecraft. Therefore, for each scatter event
we need to determine the angle of the scatter in the plane of the
instrument, as determined by the center-to-center direction of the
two detectors involved, as well as the instantaneous rotation angle
of the RHESSI spacecraft. In our plots of the ASAD,
0$^{\circ}$ is defined as east in solar coordinates,
and the azimuthal angle is measured
clockwise relative to this direction (i.e., 90$^{\circ}$ is solar north).

\subsection{ASAD Derivation}

Here we present a detailed description of our analysis method for
deriving the ASAD for each flare.
All manipulations of the raw photon data, to the point of binning the
scatter directions into an ASAD, are handled with standard RHESSI
software under the SolarSoft (SSW) system:

\noindent {\bf Step 1.} Read in all RHESSI interactions during the defined time interval.

By default, we always begin with all the interactions that occurred within
the RHESSI detectors during our time interval of interest.

\noindent {\bf Step 2.} Reject all events which do not occur among the rear segments.

By rejecting all of the other interactions, we are effectively treating
the front segments as passive shielding from low-energy photons. Thus,
we are including some photons in our ASAD which scattered first in
a front segment, effectively depolarizing the photon
before subsequently scattering between rears.
These events are properly included in our calculations
of the modulation factor (Sec.\,4.6), treated
as a background component which decreases
the measured modulation.
The same is true of all passive material in the spacecraft.

\noindent {\bf Step 3.} Apply the energy calibration.

Use standard RHESSI software to perform a careful energy
calibration and determine the energy of each interaction.

\noindent {\bf Step 4.} Reject all interactions $<$10\,keV.

These events are predominantly noise triggers in the rear detectors.

\noindent {\bf Step 5.} Determine rear-rear coincidences, reject other
interactions.

Coincidences are determined (Sec.\,4.2, Fig.\,3) by finding interactions in two
separate rears with identical
time tags ($\Delta$t = 0\,b$\mu$s), while requiring that
there were no other interactions within 4\,b$\mu$s in any
rear, either before or after this coincident
pair. All non-coincident events are rejected at this point.

\noindent {\bf Step 6.} Reject any coincident pair where either interaction
is $<$30\,keV.

This step is designed primarily to reject chance coincidences
due to higher background from flare photons at low energies.
Relatively few photons
above 0.2\,MeV will lose $<$30\,keV in Compton scatter interactions. 
In practice, no photons at these low energies can directly reach
the rear segments through the fronts, so these events are dominated
by low energy photons which have scattered off the spacecraft or Earth's
atmosphere and into the rears.

\noindent {\bf Step 7.} Reject any coincident pair whose combined ``photon'' energy
does not lie in the 0.2--1\,MeV range. 

This is the optimal photon energy band for Compton $\gamma$-ray polarimetry
with RHESSI for solar flares. Below 0.2\,MeV, few photons have enough
energy to scatter out of one detector and into another. Above 1\,MeV,
the solar flare emission is dominated by nuclear line emission.
We also reject photons within a 20-keV band centered on the 0.511-MeV
electron-positron annihilation line as well. (We present results in the
0.2--0.4\,MeV and 0.4--1\,MeV bands separately in Sec.\,4.7.4.)

\noindent {\bf Step 8.} Reject coincident pairs inconsistent with Compton scatter
kinematics.

The most reliable coincident events for studying polarization
will be those that scatter a single time in one detector, and
then are completely absorbed (in one or more interactions)
in a second detector. Since RHESSI
does not resolve the photon interactions within a detector, we
can not determine this absolutely for any coincident pair. But we
can check that the energies recorded in the two detectors are
consistent with a Compton scatter of a photon with total initial energy
given by the sum of the interactions in the two detectors. In addition,
we make a more stringent cut at this point. Since real photon scatters
between the detectors are likely large-angle Compton scatters, which
are the scatters most sensitive to polarization, we require that
the energies measured in the two detectors be consistent with a Compton scatter angle
in the range 45$^{\circ}$--135$^{\circ}$.
These cuts help reject chance coincidences and photons with incomplete
energy depositions, as well as forward-scatter and back-scatter events
that are least sensitive to polarization (but see Sec.\,4.7.3).

Since we only measure the energies in the two detectors for coincident events,
there is not enough redundant information to verify that the photon was fully absorbed,
and that the scatter angle lies within the optimal range. However, we can verify
that the energies are consistent with these conditions. To be explicit, we start
with the assumption that the photon was fully deposited, i.e. the initial photon
energy is equal to the sum of the two measured energies. Then we check both possible
interaction orderings to verify whether either one is consistent with a photon, of that
assumed initial energy, scattering between the two detectors with a a Compton scatter angle
in the range 45$^{\circ}$--135$^{\circ}$. If neither interaction order is consistent with
this Compton scatter angle range, we reject the event -- these are likely to be
photons that had Compton scatter angles outside the optimal range, or photons that 
were not full stopped in the detectors. If one or both of the potential interaction
orderings are consistent with a Compton scatter angle in this optimal range,
we assume the event is good and keep it in the analysis.

\noindent {\bf Step 9.} Determine the most-likely direction
of scatter between the two detectors, in the instrument
frame of reference (FOR).

From the same Compton kinematic analysis in Step 8, we can
determine the most-likely
order in which the photon scattered between the detectors. This
determines the photon scatter direction in the instrument FOR,
which we take to be the center-to-center direction between the
detector pairs. We note that the Compton scatter information does not
help to constrain this azimuthal scatter angle direction (we have
to take the direction as between the two detector centers); however, this
information helps determine the order in which the photon scattered
between the detectors. From our analysis in Step\,8, if only one potential
interaction ordering is consistent with the Compton kinematic cuts, we
assume that ordering is the correct one. If \textit{both}
potential interaction orderings are consistent with the Kinematic cuts, then
we assume that the interaction with the smaller energy deposition was the
initial scatter. This assumption is consistent the Compton scattering having a
higher cross section for scattering in the forward direction, and we confirmed
that this choice of interaction ordering has a higher probability from the
Monte Carlo simulations. (Note, however, that
since polarization modulations are 180$^{\circ}$-symmetric, the choice of this
interaction order should not affect the final results, as long as a consistent
choice is made for these ambiguous cases.)

\noindent {\bf Step 10.} Correct for the spacecraft rotation.

Using standard RHESSI analysis software, we determine the rotation
angle of the spacecraft (relative to fixed solar coordinates) at the instant of
each coincident event. Then the scatter direction in the instrument
FOR for each interaction is corrected to the scatter direction
relative to the solar disk.

\noindent {\bf Step 11.} Create the ASAD.

The individual scatter directions are binned to create an
ASAD for the time interval. We
bin the events in 24 bins, 15$^{\circ}$ each, covering a full 360$^{\circ}$.

Fig.\,4 shows the raw flare ASAD for 0.2--1\,MeV scatter events 
for both the 23 July 2002 flare and the 28 October 2003 flares.
These raw distributions are really the sum of three 
components: the true scattered photons from the flare, scattered photons
from the background, and chance coincidences between detectors.
In addition, there is another background component which includes
photons which first scatter in the passive spacecraft material before
subsequently scattering between the rears; however, this component
is treated separately through the modulation factor (Sec.\,4.6).
An unpolarized signal should be flat, while polarization should
manifest itself in this ASAD as a sinusoidal
component with a period of 180$^{\circ}$. Before we search for a modulation,
we need to determine the background and chance coincidence distributions, and
the modulation factor for the observations.

\subsection{Coincidence Window}

As noted above, RHESSI does not have detector-coincidence electronics;
therefore, photon scatter events between detectors have to be determined
by the interaction timing information.
In order to determine the detector coincidence timing window, as well
as the rate of chance coincidences, we perform an identical analysis
to the steps identified above, except at Step 5 we record the time
difference between every pair of interactions adjacent in the event
list. These time differences are binned by the RHESSI time resolution,
1\,b$\mu$s.
The distribution of time differences for our two flare intervals are
plotted in Fig.\,3. The real scatter coincidences (both flare and
background photons) show up as a strong
peak in this distribution at $\Delta$t = 0\,b$\mu$s. The underlying 
continuum is the chance coincidence distribution, which is expected
to be an exponential distribution with a time constant given by
the average time between interactions (or, the inverse of the interaction
rate).
This distribution allows us to determine the
coincidence window for true photon scatter events ($\Delta$t = 0\,b$\mu$s)
and the underlying rate of chance
coincidences within this window. We note that the number of events in
the coincidence window for both data sets is very close to the number
of events in the ASAD derived through the steps
above. (Given the slight differences required in deriving the curves,
we expect the peak in the time-difference distribution to be slightly
lower.)

\subsection{Chance Coincidences}

In order to subtract the chance-coincidence ASAD
from the raw distribution shown in Fig.\,4, we follow the same steps as
outlined above, except instead of requiring the interactions to occur
with $\Delta$t = 0\,b$\mu$s, we require the time difference between the
interactions to be $\Delta$t = 4\,b$\mu$s. This creates a scatter angle
distribution of interactions that are purely chance coincidences, with
the number of chance coincidences nearly identical to the number within our
raw flare ASAD. The chance coincidence ASADs are shown
in Fig.\,5. 

For unpolarized signals, we expect all the ASADs to be flat.
A polarized component will show up as a sinusoidal modulation on top
of this flat distribution with a period of 180$^{\circ}$, since Compton
scattering is symmetric around the polarization axis. We do not
expect a sinusoidal component in this distribution with a 360$^{\circ}$
period; however, these components show up quite clearly in both the
raw flare ASADs (Fig.\,4) and the chance-coincidence ASADs
(Fig.\,5). Indeed, the 360$^{\circ}$ component appears to be an instrumental
effect due solely to
the chance coincidences, and disappears from the flare ASAD
when we subtract off the chance-coincidence component. We do not
understand the origin of this 360$^{\circ}$ period component, but it appears
to be a systematic instrumental effect arising from the chance coincidences,
and subtracts away with this component.

In Sec.\,4.7.3 we discuss how we used these chance coincidence ASADs to 
verify that the chance coincidences are consistent with zero polarization,
as expected.

\subsection{Background Scatters}

Since the RHESSI detectors are unshielded, they are subject to a constant
background of true cosmic and atmospheric photons, some fraction of which
scatter between the detectors identically to the flare photons. To determine
the background component to our scatter angle distribution, we identify
two background time periods for each flare identical in length to our
flare time intervals: one background period during the orbit before the
flare, and one interval during the orbit following the flare. These time
intervals are defined in Table\,1. The average background ASADs are shown in
Fig.\,6, scaled by by the scatter livetimes discussed in Sec.\,4.5.

The difference in the scatter-coincidence rates for the two background
periods we chose for each flare are larger than anticipated from purely statistical
fluctuations, which is expected since the RHESSI background varies significantly
on orbital timescales. We characterize our systematic uncertainty on the background
scatter rate as half of the difference between the two backgrounds selected for
each flare. This systematic uncertainty dominates the overall uncertainty on our
average background-subtracted scatter rate during the flare.

The average background ASADs show no significant signs
of the 360$^{\circ}$ period components, likely due to the small number of
chance coincidences during the background intervals. The background
distributions also do not show any significant 180$^{\circ}$ modulations.
In Sec.\,4.7.3 we discuss how we used these average background ASADs to 
verify that the background events are consistent with zero polarization,
as expected.

\subsection{Scatter Livetime}

In Figs.\,2a \& 2b we showed the 0.2--1\,MeV single interaction rates in the
rears. In Figs.\,2c \& 2d we show the total
coincidence rates among the rears for all the
detector pairs shown in Fig.\,1, and the chance coincidence rates among
the same pairs as determined by the method in Sec.\,4.3.
As can be seen in Figs.\,2c \& 2d, the chance coincidence rates
do not scale linearly with the single interaction rates. Indeed, we see that
when the rates get high enough the coincident events are dominated by
the chance coincidences, with the number of real scatter events dropping.
The system responds as if there is an {\it effective livetime} to real
scatter events, which drops to zero if the event rate is high enough.

We note that this is a separate problem from determining the ratio of
real scatter events to chance coincidence events, which is straightforward
to determine from the peak in Fig.\,3 at $\Delta$t = 0\,b$\mu$s
relative to the underlying continuum.

This effective scatter livetime and saturation at high count rates
affects our analysis in two ways. First, in order to maximize our
real scatter signal and limit the chance coincidence background, we
select only time intervals where the photon scatter rate is greater
than the chance coincidence rate, as determined by making time-resolved
versions of Fig.\,3. This criteria did not affect the
2002 July 23 analysis, but explains why our 2003 October 28 flare
analysis window is offset several minutes after the peak single
interaction count rate in the rears.

The second effect on the analysis is determination of this effective
scatter livetime. Since we derive the background ASADs from off-flare
time intervals, direct subtraction of the average background ASAD will
{\it over subtract} background from the flare ASAD since the real 
background scatter rate is reduced by the effective scatter livetime
during the time of the flare. Therefore, to properly subtract this 
component our average background ASADs must be
corrected by this effective scatter livetime before subtraction from
the raw flare ASADs.
(The average background ASADs shown in Fig.\,6 have already
been corrected by these livetime factors.)

For our flare periods, estimating the
scatter livetime is straightforward if we make the reasonable
assumption that background photon scatters and flare photon scatters
are suppressed by the same effective livetime during the flare.
These scatter livetimes were determined
by plotting the ratio of true scatter events (determined by time-resolved
versions of Fig.\,3) to single events, as a function of the overall singles
event rate (Fig.\,7). If there were no effective deadtime to scatter events,
we would expect these plots to be flat; however, we can clearly see a roll-off in
this ratio at high count rates. By determining the average ratio at low count
rates, and the average ratio for our observation period, we can take their
ratio as our effective scatter livetime (Fig.\,7). For our analysis
periods listed in Table\,1, we estimate the fractional scatter livetimes
to be 0.73$\pm$0.01 for the 2002 July 23 flare, and 0.74$\pm$0.01
for the 2003 October 28 flare.

\subsection{Modulation Factor}

Before we can determine the intrinsic polarization of the flares, we need to
estimate the modulation factor for the RHESSI instrument.
The modulation factor will depend on a number of factors including the
detector geometries, the
spectral shape in the 0.2--1\,MeV range, and the energy and Compton kinematics
cuts that we perform in deriving the ASAD.

Photons which first scatter in the passive material of the spacecraft
will be effectively depolarized before subsequently scattering between rears.
While these are effectively a background component in the flare ASAD, they
are not a component that we can identify and subtract as a background. Instead,
these events must be properly included in our calculations
of the modulation factor, treated
as a background component which will decrease measured modulations,
and therefore the modulation factor. These events have been properly included
and accounted for in our calculations of the modulation factor.

In order to determine this modulation factor, we utilized a detailed RHESSI
mass model, implemented for the MGEANT interface of the CERN GEANT Monte
Carlo package \citep{stu00}, with the GLEPS package which includes
$\gamma$-ray polarization \citep{mcc02}.
This mass model was developed to study the RHESSI background
components \citep{wun04}. As an input spectrum, we chose
the best-fit spectral distribution determined for the 23 July 2002 flare,
which corresponds to a broken power law with an index break from 2.77 to
2.23 at 617\,keV \citep{smi03}.
We verified that this modulation factor does not
significantly change for pure power law spectra with spectral indices
ranging between 2 and 3, which encompasses our broken power law spectrum.
We performed the same data cuts on the simulated interactions that we performed
on the real interactions. By simulating 100\% polarized photons, we can measure
the modulation factor directly from the simulated ASAD, shown in Fig.\,8, to
derive $\mu_m$ = 0.32$\pm$0.03.
We verified the results of this Monte Carlo by using MGEANT without the
the polarization package, and kept track of the modulation factor for each
individual scatter event, taking into account whether a photon first scattered
in a front segment. This semi-analytical approach yielded an estimate of
$\mu_m$ = 0.33$\pm$0.01, in agreement with the more detailed simulations.

\subsection{Results}

At this point we have three ASADs for each flare:
the raw flare ASAD, the chance-coincidence ASAD,
and the average-background ASAD. We subtracted the
chance-coincidence ASAD and the
average of the two background ASADs 
(scaled by the scatter livetimes) from the raw flare
ASAD. This produces
the residual flare ASADs for the flare photons alone, which
are shown in Fig.\,9.

For the two residual background-subtracted flare ASADs,
we can now search for significant modulations corresponding to 180$^{\circ}$
periods. For each distribution, we fit a function of the form of Eqn.\,2
to determine the amplitude of the potential modulated component. 
Correcting for the modulation factor we can determine the
intrinsic polarization fraction for the $\gamma$-ray photons from these
two solar flares.
In addition, we compare the modulation fit to unpolarized distributions
to determine the significance of the measured signals.

\subsubsection{23 July 2002}
 
Fig.\,9a shows the best-fit modulation curve to the 23 July 2002 flare
scatter angle distribution. For this choice of binning, the amplitude
of the modulation component is 19$\pm$8, with a 2.4$\sigma$ significance.
The average is 277$\pm$10, with the uncertainty dominated by the
systematic uncertainty in the background level during the flare.
The ratio of these two yields the
polarization of the incident flare $\gamma$-rays multiplied by the
instrumental modulation
factor. For this flare, $\mu_m \Pi_s$ = 0.069$\pm$0.029, the uncertainty
of which includes the fact that the polarization direction is not known
\textit{a priori}. In order to
determine the significance of this modulation, we performed a simple
Monte Carlo simulation assuming an unpolarized (flat)
distribution, and the average measured
uncertainty, to determine how often we would randomly fit a modulation
of this amplitude for an unpolarized source. Given our measurement
uncertainties, the chance probability of fitting a modulation
of this amplitude is 7.7\%.

The minimum in the modulation curve corresponds
to a direction of the polarization vector of
$\eta$ = 78$^{\circ}$ $\pm$ 13$^{\circ}$ (north of east). 
Projected onto the solar disk (Fig.\,10), this
polarization vector is perpendicular (within its uncertainty)
to the direction from the disk center toward the solar flare,
which corresponds to a positive polarization by convention.

Correcting for the modulation factor, the fractional linear polarization
for this flare is $\Pi_s$ = 0.21 $\pm$ 0.09 in the 0.2--1\,MeV band.
(The average measured photon energy over this band is 0.45 MeV.)
For comparison, in Fig.\,9a we have plotted the modulation level
for a 100\% polarized signal. While our absolute detection is at the
marginal 2.4$\sigma$ level, we are still strongly constraining the polarization
level of this flare. At the 99\% confidence level (3$\sigma$), we
constrain this polarization to lie within the range -6\% to +48\%.

\subsubsection{28 October 2003}

Fig.\,9b shows the best-fit modulation curve to the 28 October 2003 flare
scatter angle distribution. The amplitude
of the modulation component is 24$\pm$12, with a 2.0$\sigma$ significance.
The flat background is 685$\pm$51, with the uncertainty once again dominated by the
systematic uncertainty in the background level during the flare.
The ratio for this flare yields $\mu_m \Pi_s$ = 0.035$\pm$0.018.
For our measurement uncertainties, the chance probability of fitting
a modulation of this amplitude is 14\%.

The minimum in the modulation curve corresponds
to a direction of the polarization vector of
$\eta$ = 101$^{\circ}$ $\pm$ 15$^{\circ}$ (north of east). This
polarization vector is parallel to the direction from the
disk center toward the solar flare (Fig.\,10), corresponding to
a negative polarization by convention. 

Correcting for the modulation factor,
the fractional linear polarization for this flare
is $\Pi_s$ = -0.11 $\pm$ 0.05 in the 0.2--1\,MeV band.
(The average measured photon energy over this band is 0.50 MeV.)
Once again, we have also plotted in Fig.\,9b the
modulation amplitude for a 100\% polarized signal, demonstrating
that we are strongly constraining the polarization level for this
flare. At the 99\% confidence level (3$\sigma$), we
constrain this polarization to lie within the range -26\% to +4\%.

\subsubsection{Null Results}

As a critical test of our polarization techniques and characterization of
potential systematic errors,
we need to verify that signals we expect to be unpolarized do not show significant
modulations. These signals include both our chance coincidence backgrounds and our
true scattered-photons backgrounds, neither of which we would assume \textit{a prior}
to be modulated.

For the background scatters in Fig.\,6, we determined the best-fit modulation amplitudes to be
3 $\pm$ 2 for the 23 July 2002 flare, and 5 $\pm$ 3 for the 28 October 2003 flare. Correcting
for the modulation factor, these modulations correspond to effective
polarizations of 5 $\pm$ 4\% and 4 $\pm$ 3\% for the two flare backgrounds, respectively.
These results demonstrates that our background scatters
are not polarized, and that systematic modulations are restricted for real photon-scatter
events in our analysis below the 5\% polarization level.

For the chance coincidence events in Fig.\,5, we also determined the best-fit modulation amplitudes
to be 2 $\pm$ 3 for the 23 July 2002 flare, and 10 $\pm$ 5 for the 28 October 2003 flare. (Note the
fits to the 180$^{\circ}$ modulations are not significantly affected by the orthogonal 360$^{\circ}$
components.) Once again correcting for the modulation factor, the effective polarizations
corresponding to these modulations for the chance coincidences alone are 4 $\pm$ 6\%  and 12 $\pm$ 6\%
for the two flare periods, respectively. However, when we compare the absolute value of these best-fit
modulations with those measured for the background-subtracted flare ASADs above, we can see
that the potential effects of these chance-coincidence modulations on the overall measured flare photon
polarizations are 2 $\pm$ 3\% and 4 $\pm$ 2\% for the two flares respectively. Therefore, we can also
limit the effects of systematically-induced modulations from the chance-coincidence events to below
the 5\% polarization level. 

As a further check of null results, we repeated the polarization analysis for
large-angle \textit{backscatter} events in the
instrument (Fig.\,11). For the analysis above (ultimately deriving Fig.\,9) we only accepted
scatter events which were consistent with Compton scatter angles in the range 45$^{\circ}$--135$^{\circ}$
(Step\,8 of our analysis method). This range of Compton scatter angles is the most
sensitive to polarization modulations, hence this cut helped maximize the signal-to-noise. Small-angle
scatters (0$^{\circ}$--45$^{\circ}$) and large-angle backscatters (135$^{\circ}$--180$^{\circ}$) are less sensitive to
the incoming photon polarization, and therefore with our marginal sensitivities we expect null results for
modulation curves in these cases. For our data cuts, small-angle scatters are dominated by chance coincidence
events (due to favorable low-energy coincidences), while for backscatter events the chance coincidences are 
nearly negigible. Since we have already limited the systematic polarizations for chance coincidences, we
repeated the full analysis for backscatter events to verify null results.
For Steps\,8 \& 9, we modified the Compton kinematic cut criteria so that if either
of the potential interaction orderings
were consistent with a Compton scatter angle in the range 135$^{\circ}$--180$^{\circ}$, then the event
was accepted for further analysis.
The measured modulations for these backscatter
events, shown in Fig.\,11, are $\mu_m \Pi_s$ = 0.06$\pm$0.06 for the 2002 July 23 flare,
and $\mu_m \Pi_s$ = 0.03$\pm$0.04 for the 28 October 2003 flare, consistent with our null hypothesis.

Based on these results, we are confident that we have limited the systematic uncertainties in these
measurement techniques to below the 5\% polarization level.

\subsubsection{Energy Bands}

We have presented our polarization analysis and results for the entire 0.2-1\,MeV band. For completeness
we have also performed the identical analysis over the 0.2-0.4\,MeV band and the 0.4-1\,MeV band for
comparison. (Including calculation of the modulation factors for each of these smaller energy bands.)
The final results of these analyses are presented in Table\,2. 

At first glance, the results presented in Table\,2 suggest that the 2002 July 23 flare is more strongly
polarized in the 0.2-0.4\,MeV band than the 0.4-1\,MeV band, while the 2003 October 2002 flare is
more strongly polarized in the higher energy band. We stress that with our limited statistics these
trends should be considered with caution.

\section{Discussion}

Despite the marginal detection of polarization for both of these
flares, together they exhibit a few interesting properties.

Both of the flares show physically significant directions for their
polarization vectors (Fig.\,10), either perpendicular or parallel to
the flare--disk center direction. While these detections are both
marginally significant, the alignment of the polarization
vectors along the two physically-possible directions supports
the presumption that these are not spurious detections. It is
especially intriguing that the polarization vector appears to
have rotated by 90$^{\circ}$ from the central disk toward
the limb. If confirmed by further events, this trend will place
strong constraints on the underlying beamed electrons.

Our measurements are consistent with the general expectation
that the level of polarization will increase with increasing viewing
angle, regardless of the underlying beamed electron distribution.
For flares near the disk center (small viewing angles), the
symmetry of the geometry alone should require the polarization level
to approach 0\%.

The levels of polarization measured are generally consistent with theoretical
predictions for beamed electron distributions ($\leq$25\%),
and marginally inconsistent with predictions for isotropic distributions
($<$10\%). In Fig.\,12 we have plotted our measured polarizations
along with one theoretical prediction for several $\gamma$-ray
photon energies for a beamed
electron model assuming a half opening angle of 30$^{\circ}$ and
and accelerated electron spectrum of $E^{-3.5}$ \citep{bai78}.
Our results are generally consistent with this model, but we would
clearly like to perform this analysis with several energy bands
and many more flares -- which is currently beyond the
serendipitous RHESSI sensitivity to $\gamma$-ray polarization which we
are exploiting.

We plan to continue analyses of these and other RHESSI solar flares in
hopes of better constraining the $\gamma$-ray polarization as a function
of both viewing angle and photon energy. These studies can clearly play an
important role in illuminating the underlying electron beaming, and particle
acceleration mechanisms, in solar flares. This paper has attempted to
achieve two milestones in this endeavor: to verify the sensitivity of
RHESSI to solar $\gamma$-ray polarization, and to establish the
overall level of $\gamma$-ray polarization we can expect to observe for
solar observations.

  \acknowledgments
  The authors are grateful to H. Hudson for comments on this manuscript, and
  C. Wunderer for providing the RHESSI MGEANT mass model.
  SB and WC are grateful for support under NASA and California Space Institute.

  \clearpage

  \begin{figure}
  \epsscale{0.60}
  \plotone{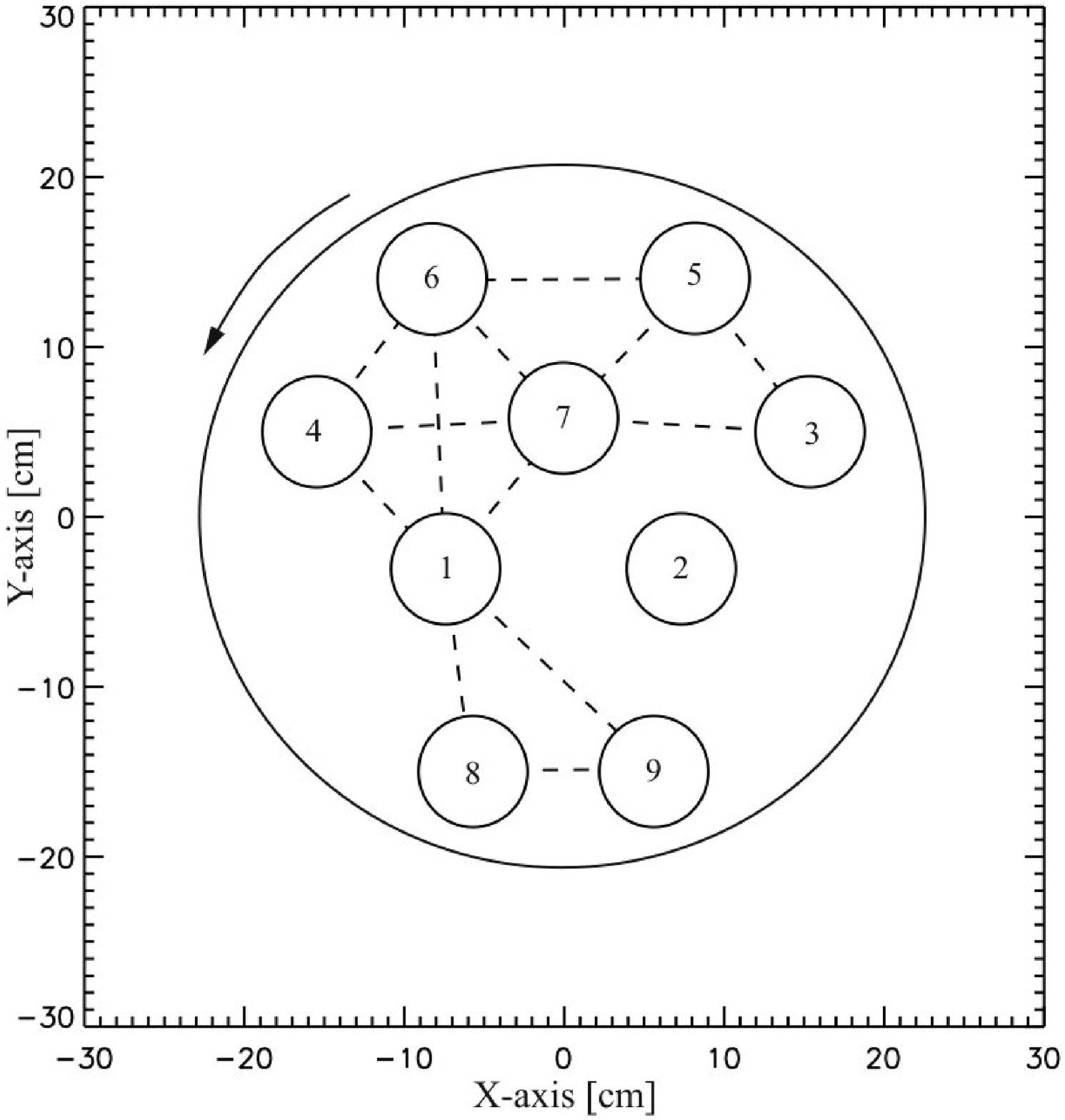}
  \caption{Configuration of the nine RHESSI detectors in the plane of the
   spectrometer as viewed from the front of the spacecraft, which rotates
   in the direction indicated with a 4-s period.
   The dashed lines represent detector-detector coincidence
   paths among the neighboring rear segments used in this analysis. Detector
   \#2 was not used (see text). \label{fig1}}
  \end{figure}

  \clearpage

  \begin{figure}
  \epsscale{1.0}
  \plotone{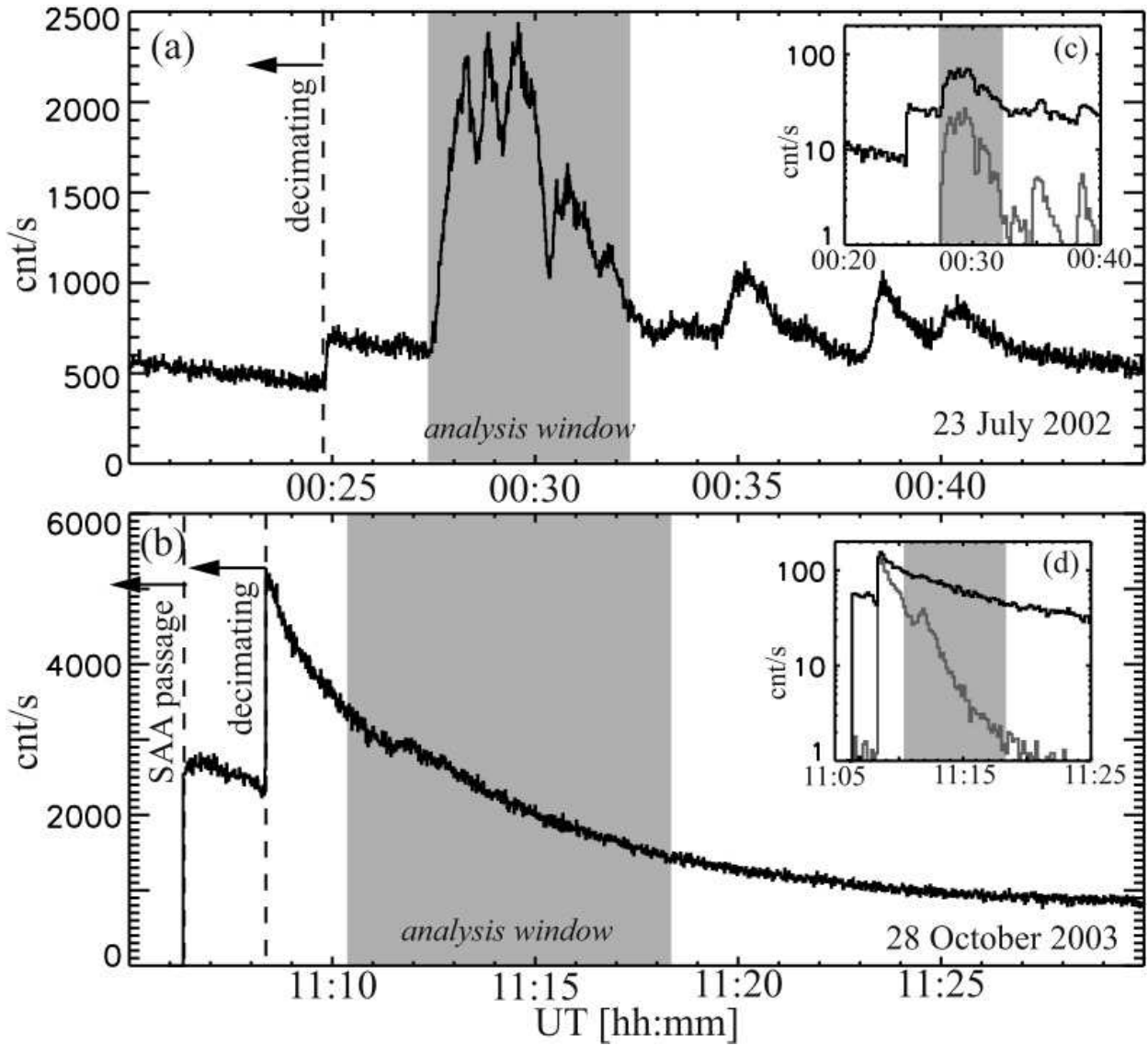}
  \caption{Lightcurves of 0.2-1\,MeV {\it single} events in the
   RHESSI rear segments for (a) the 23 July 2002 flare, and (b) the
   28 October 2003 flare. The insets (c) and (d) show the total coincidence
   rates (dark line) and the chance coincidence rates (light
   line) among the rears during the flare. The periods we used in this analysis
   are shown in grey. \label{fig2}}
  \end{figure}

 \clearpage

  \begin{figure}
  \epsscale{.70}
  \plotone{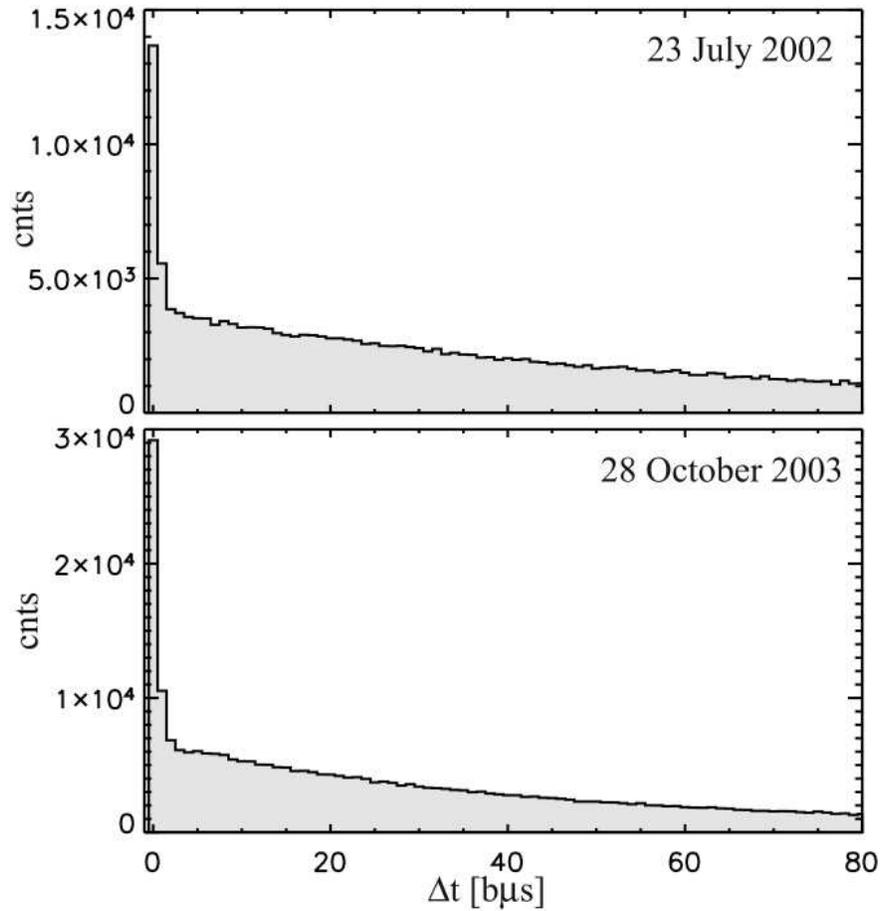}
  \caption{Distribution of waiting times ($\Delta$t)
   between single interactions for each flare. The peak at $\Delta$t = 0\,b$\mu$s is
   a clear signal of the true photon-scatter coincidences. The underlying
   continuum is due to chance coincidences. There are some true scatter
   coincidences in the $\Delta$t = 1\,b$\mu$s bin, but we have excluded
   them to maximize the signal-to-noise. \label{fig3}}
  \end{figure}

 \clearpage

  \begin{figure}
  \epsscale{.60}
  \plotone{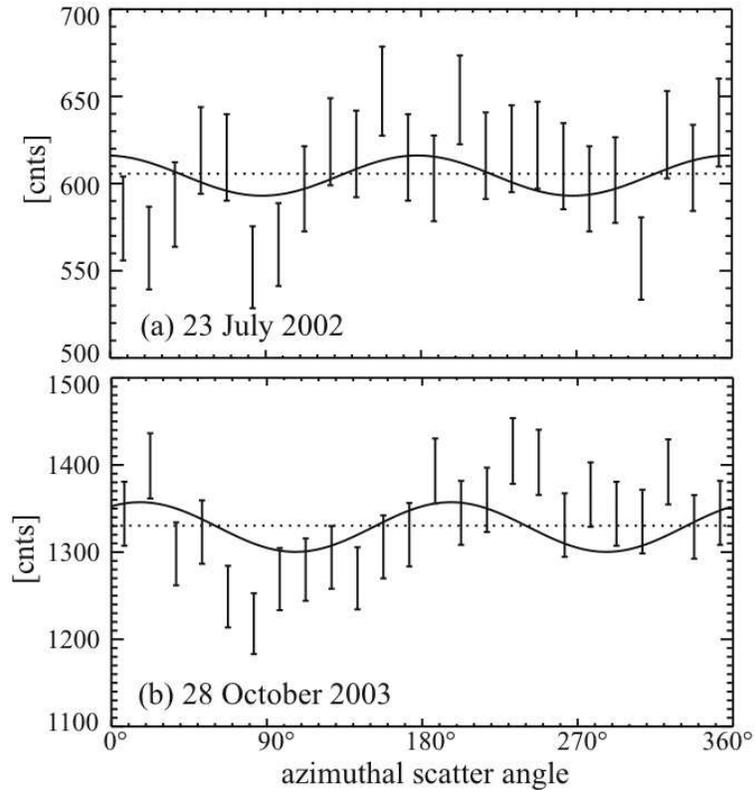}
  \caption{The 0.2-1\,MeV raw flare ASADs (no background subtracted) for
  (a) the 2002 July 23 flare,
  and (b) the 2003 October 28 flare. Averages are shown for each ASAD
  (dotted line). \label{fig4}}
  \end{figure}

 \clearpage

  \begin{figure}
  \epsscale{.60}
  \plotone{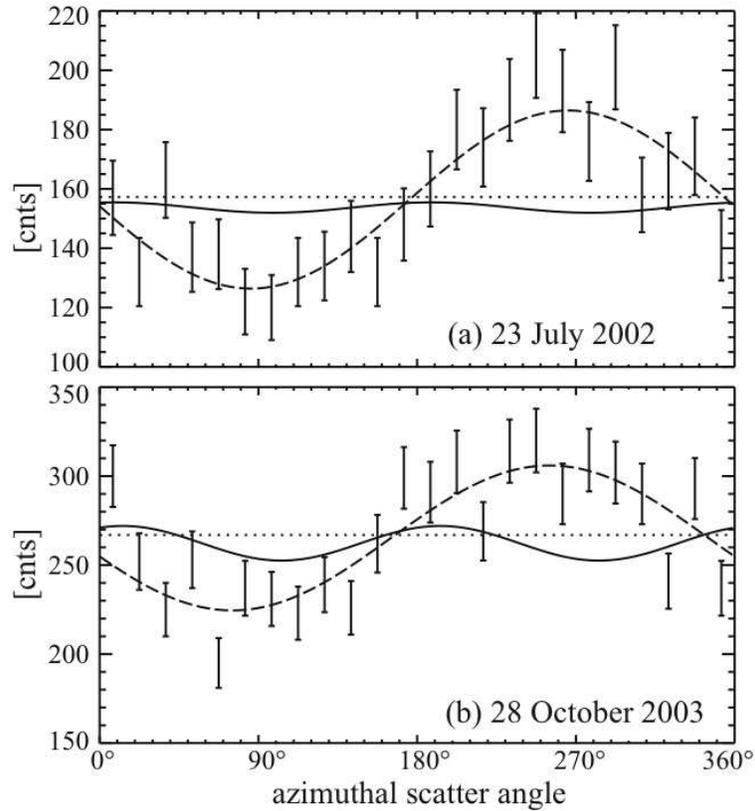}
 \caption{The 0.2-1\,MeV chance-coincidence ASADs ($\Delta$t = 4\,b$\mu$s) for
  (a) the 2002 July 23 flare,
  and (b) the 2003 October 28 flare. Averages are shown for each ASAD
  (dotted line), as well as the best-fit 360$^{\circ}$ sinusoidal fits (dashed line),
   and the best-fit modulations (solid line). The 360$^{\circ}$ component
   appears to be a systematic instrumental effect associated with these chance
   coincidences, and subtracts away in our final flare ASAD. \label{fig5}}
  \end{figure}

 \clearpage

  \begin{figure}
  \epsscale{.60}
  \plotone{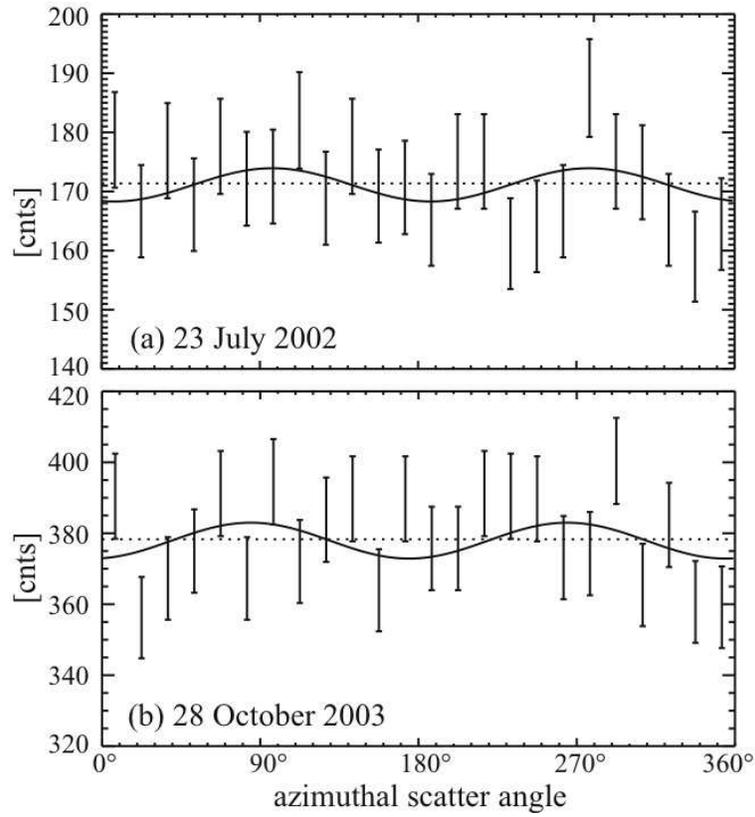}
  \caption{The 0.2-1\,MeV average background ASADs (taken during orbits immediately
  before and after the flares) for
  (a) the 2002 July 23 flare,
  and (b) the 2003 October 28 flare. Averages are shown for each ASAD
  (dotted line), as well as the best-fit modulations (solid line).\label{fig6}}
  \end{figure}

 \clearpage

  \begin{figure}
  \epsscale{.60}
  \plotone{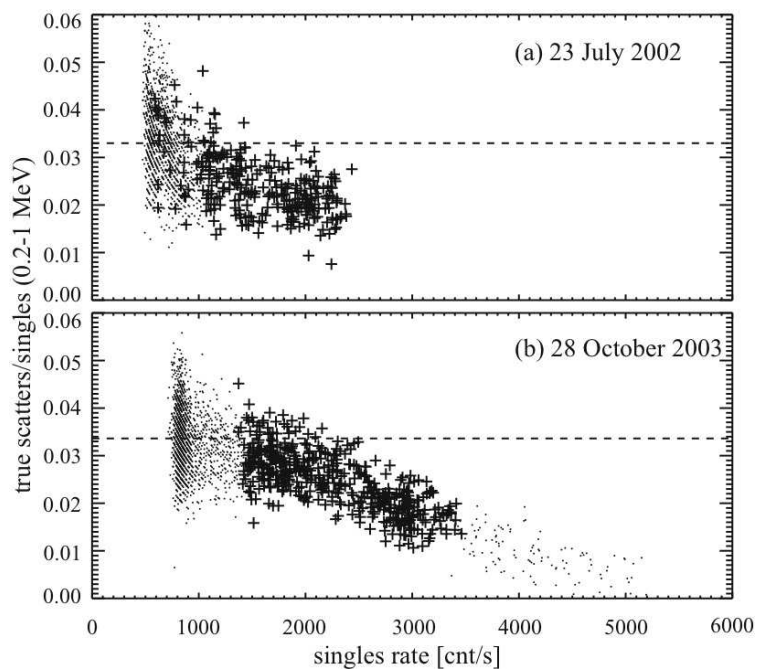}
  \caption{The ratio of 0.2-1\,MeV true photon scatters to singles count rate, as a function
  of the singes count rate. The dots show all data around the time of the flare, while the crosses
  are from the times taken for the polarization analysis. For a constant scatter livetime we would
  expect this curve to be flat -- the drop at higher count rates show a corresponding drop in the
  effective scatter livetime. By comparing the average of this ratio for times used in our analysis
  (crosses) to the average at low count rates (dashed line) we can directly measure the effective
  scatter livetimes for these analysis periods. \label{fig7}}
  \end{figure}

 \clearpage

  \begin{figure}
  \epsscale{0.70}
  \plotone{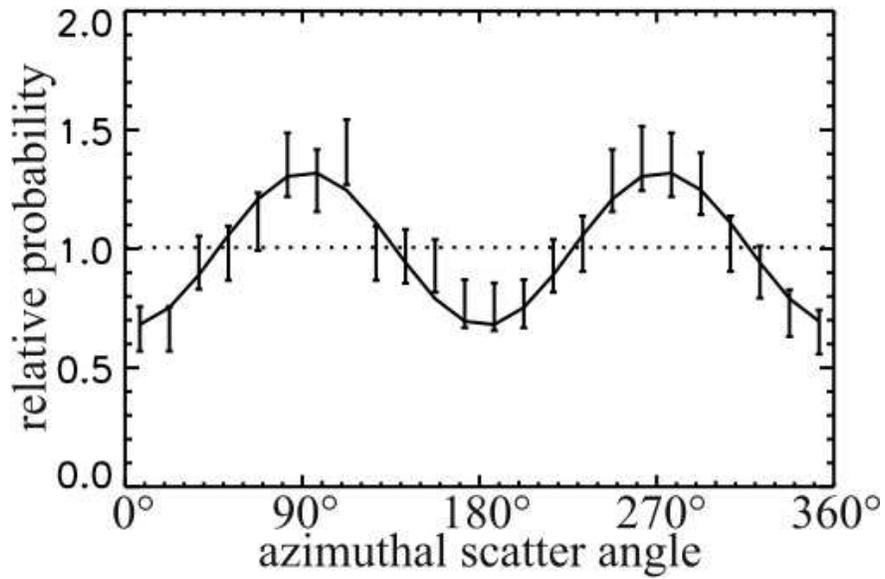}
  \caption{The simulated ASAD for 100\% polarized solar
  photons in the 0.2--1\,MeV range, assuming an input spectrum of the form measured
  for the 23 July 2002 flare (Sec.\,4.6). The modulation on this
  distribution corresponds to an instrumental modulation factor for RHESSI
  of $\mu_m$ = 0.32$\pm$0.03. \label{fig8}}
  \end{figure}

 \clearpage

  \begin{figure}
  \epsscale{0.70}
  \plotone{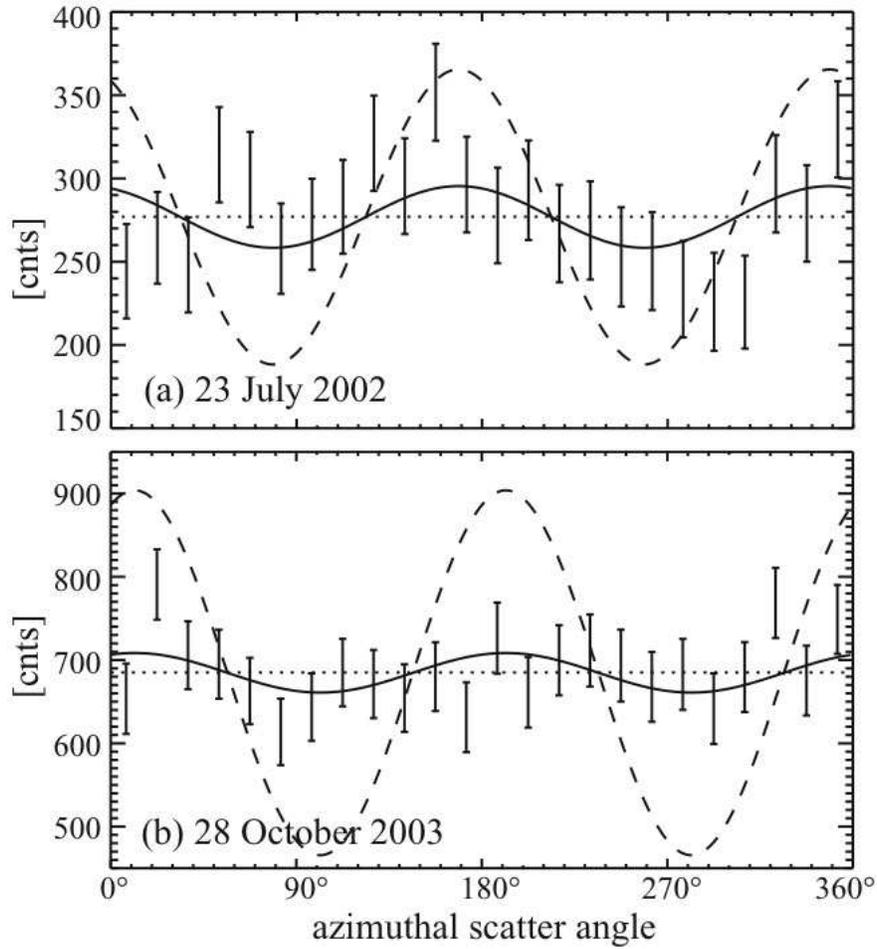}
  \caption{The 0.2-1\,MeV background-subtracted ASADs for the 2002 July 23
  flare (top), and the 2003 October 28 flare (bottom). Shown for
  comparison are the best-fit modulation (solid line), and the
  expected modulation for both unpolarized photons (dotted line)
  and 100\% polarized photons (dashed line). \label{fig9}}
  \end{figure}

 \clearpage

  \begin{figure}
  \epsscale{0.70}
  \plotone{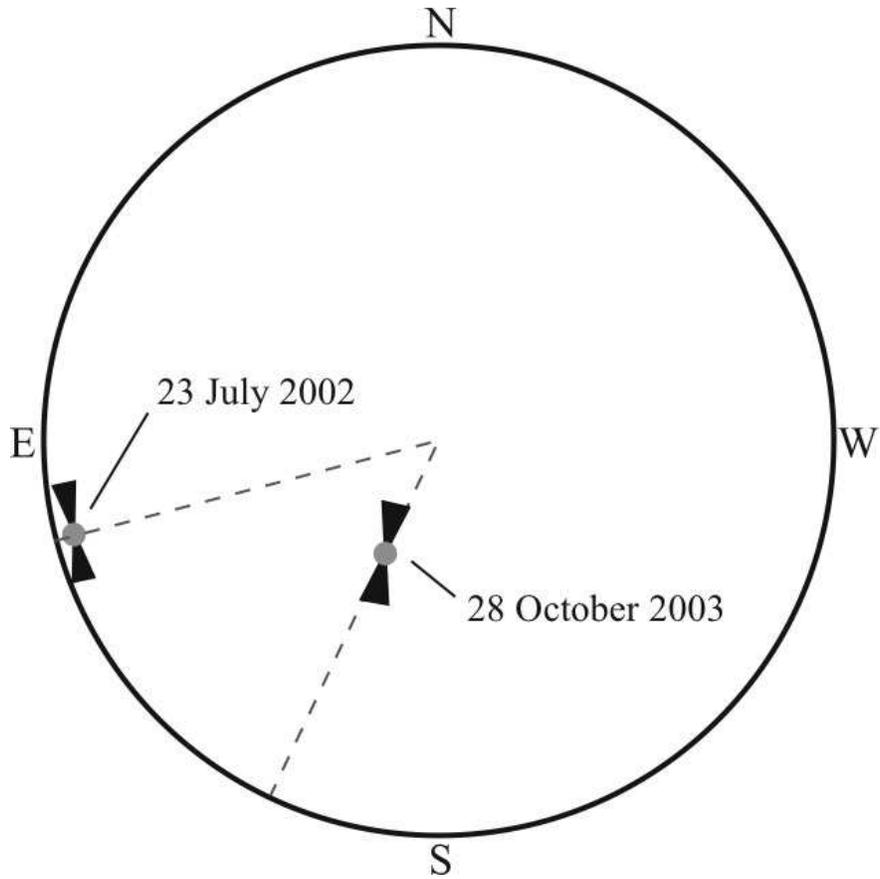}
  \caption{Diagram of the solar disk, showing the location of the
  two flares studied in this paper (grey circles). The $\pm 1\sigma$ limits on the
  0.2-1\,MeV polarization directions (black lines) are shown, as well as the radial direction
  from the disk center to each flare for reference. \label{fig11}}
  \end{figure}

 \clearpage

  \begin{figure}
  \epsscale{0.70}
  \plotone{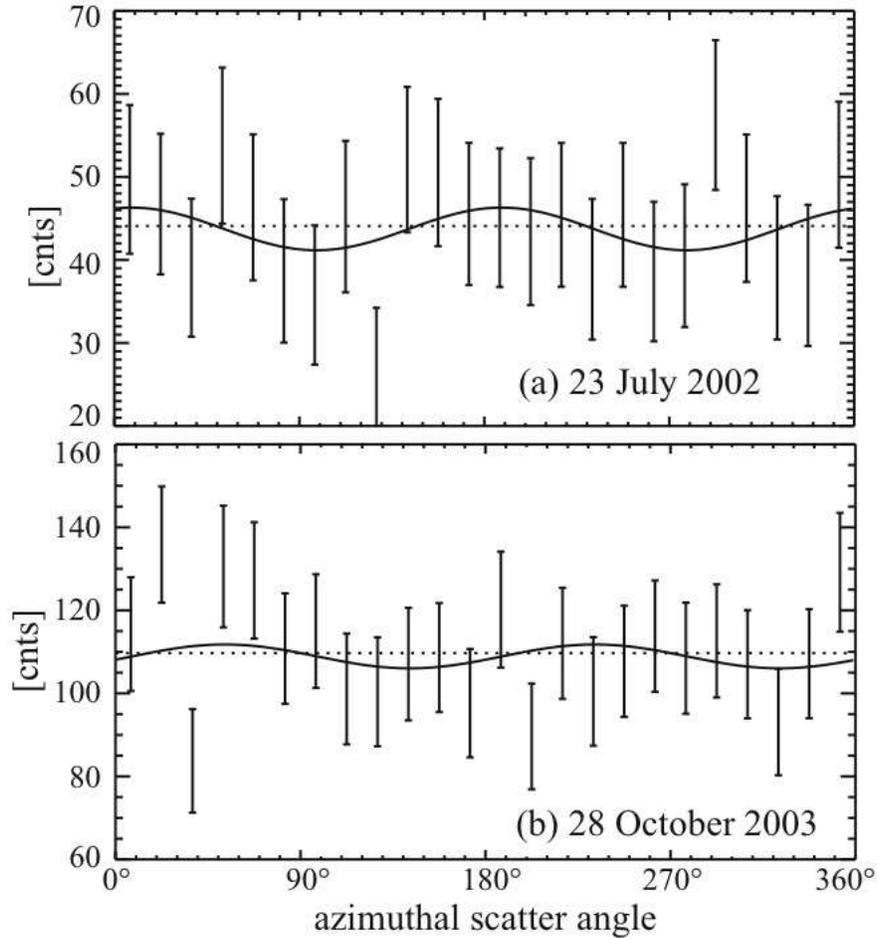}
  \caption{The 0.2-1\,MeV background-subtracted ASADs for the two flares, using
  \textit{only backscatter events} (Sec.\,4.7.3). Shown for
  comparison are the best-fit modulation (solid line),
  and the average (dotted line).  Since backscatter events are
  less sensitive to polarization than events chosen in Step\,8,
  modulations on these ASADs should not be significant. \label{fig10}}
  \end{figure}

 \clearpage

  \begin{figure}
  \epsscale{0.70}
  \plotone{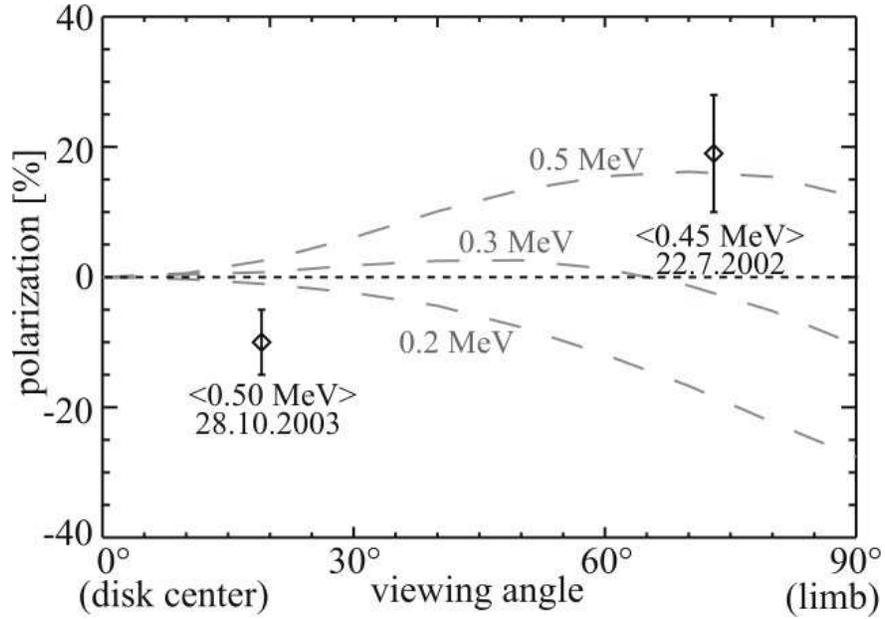}
  \caption{Our 0.2-1\,MeV polarization measurements compared with
  a theoretical model from \cite{bai78} for an accelerated electron
  spectrum of E$^{-3.5}$, beamed with into a 30$^{\circ}$
  half opening angle.
  The predictions are shown for three different photon
  energies: 0.2, 0.3, 0.5 MeV (dashed lines). The average photon energy for
  each measurement is shown for comparison. Negative polarizations
  correspond to polarization vectors aligned toward the center of the solar
  disk, while positive polarizations are perpendicular to this direction. \label{fig12}}
  \end{figure}

\clearpage
 
\begin{table}
\begin{center}
\caption{Observation time intervals.\label{tbl-1}}

\begin{tabular}{llccc}
\tableline\tableline
Flare & Observation & Date & Start & End \\
\tableline
2002 July 23 & Flare & 2002.07.23 & 00:27:20 UT & 00:32:20 UT \\
 & Background 1 & 2002.07.22 & 23:00:20 UT & 23:05:20 UT \\
 & Background 2 & 2002.07.23 & 02:10:20 UT & 02:15:20 UT \\
\tableline
2003 October 28 & Flare & 2003.10.28 & 11:10:22 UT & 11:18:22 UT \\
 & Background 1 & 2003.10.28 & 09:42:22 UT & 09:50:22 UT \\
 & Background 2 & 2003.10.28 & 11:37:22 UT & 11:45:22 UT \\
\tableline
\end{tabular}

\end{center}
\end{table}

\clearpage
 
\begin{table}
\begin{center}
\caption{Measured polarization amplitudes and directions.\label{tbl-2}}

\begin{tabular}{llccc}
\tableline\tableline
Flare & Energy Range [MeV] & $\Pi_s$ & $\eta$  \\
\tableline
2002 July 23 & 0.2--1 & 0.21 $\pm$ 0.09 & 78$^{\circ}$ $\pm$ 13$^{\circ}$  \\
 & 0.2--0.4 & 0.26 $\pm$ 0.12 & 65$^{\circ}$ $\pm$ 13$^{\circ}$  \\
 & 0.4--1 & 0.17 $\pm$ 0.15 & --  \\
\tableline
2003 October 28 & 0.2--1 & -0.11 $\pm$ 0.05 & 101$^{\circ}$ $\pm$ 15$^{\circ}$  \\
 & 0.2--0.4 & 0.07 $\pm$ 0.07 & --\\
 & 0.4--1 & -0.25 $\pm$ 0.09 & 113$^{\circ}$ $\pm$ 10$^{\circ}$ \\
\tableline
\end{tabular}

\end{center}
\end{table}

  \end{document}